# PRECISE TEST OF INTERNAL-CONVERSION THEORY:

# $\alpha_K$ MEASUREMENTS FOR TRANSITIONS IN NINE NUCLEI SPANNING $45 \leq Z \leq 78$

J.C. Hardy[1]*, N. Nica[1], V.E. Iacob[1] and M.B. Trzhaskovskaya[2]

[1]*Cyclotron Institute, Texas A&M University, College Station, TX 77845-3366, USA*

[2]*Petersburg Nuclear Physics Institute, Gatchina 188300, Russia*

**Abstract**

We have been testing the theory used to calculate internal-conversion coefficients (ICCs) by making a series of measurements of $\alpha_K$ values with precision better than ±2%. So far we have measured *E*3 transitions in three nuclei, $^{103}$Rh, $^{111}$Cd and $^{134}$Cs; and *M*4 transitions in six nuclei, $^{119}$Sn, $^{125}$Te, $^{127}$Te, $^{137}$Ba, $^{193}$Ir and $^{197}$Pt. Together, these span a wide range of *A* and *Z* values. In all cases, the results strongly favor Dirac-Fock calculations in which the final-state electron wave function has been computed in an atomic field that includes the vacancy created by the internal-conversion process.

## 1. Introduction

Except for the very lightest nuclei, where internal conversion is weakest, most nuclear decay schemes depend upon calculated internal conversion coefficients (ICCs). Electromagnetic decay intensities are usually determined from gamma-ray measurements combined with calculated ICCs. Consequently, the reliability of the calculations is a matter of some importance, especially where precise decay-scheme data are required, for example in detector calibration. Until quite recently, although various tables of calculated ICCs were readily available, most ICC measurements were relatively imprecise, being aimed only at determining transition multipolarities. Rarely were they precise enough to distinguish among the different calculations or indeed to establish if any of the calculations reproduced reality. We are rectifying this deficiency.

When we began our program of precise measurements in 2004, the then-current survey of world data (Raman *et al.* 2002) included barely twenty ICC values measured to ±2% or better, and eighty more with up to 5% precision. They were divided 45-55 between *K*-shell ICCs ($\alpha_K$) and total ICCs ($\alpha_T$), respectively. Based on these data, the authors concluded that all previous tables of ICCs exhibited a 3% systematic bias, but that a table by Band *et al.* (2002), which was new at the time, agreed well with the data (within ~1%). This new table was calculated in the framework of the Dirac-Fock method, with the exchange between bound electrons as well as between bound and free electrons treated exactly, an important improvement. Unfortunately, however, the best agreement with the available experimental data was achieved with a version of this calculation in which the final-state electron wave-function was computed in a field that did not include any provision for the atomic vacancy created by the conversion process. Yet the vacancy must be there, since it can easily be shown (*e.g.* Hardy *et al.* 2014) that atomic-shell-vacancy lifetimes are generally much longer than the time for a conversion electron to leave the vicinity of the atom. This is an unsatisfactory paradox!

We found ourselves uniquely positioned to potentially resolve the paradox. For a completely different purpose, we had efficiency calibrated an HPGe detector to high precision over a wide range of energies. This would allow us to measure the *K* x rays and γ rays from a converted transition in the same well-calibrated detector, thus affording access to the transition's $\alpha_K$ value with a minimum of systematic uncertainty. Since then, we have published $\alpha_K$ values for *E*3 transitions in two nuclei, $^{111}$Cd and $^{134}$Cs, and *M*4 transitions in six nuclei, $^{119}$Sn, $^{125}$Te, $^{127}$Te, $^{137}$Ba, $^{193}$Ir and $^{197}$Pt. These results range in precision from 0.55% to 1.9%. In addition, we have recently made a measurement of a third *E*3 transition, in $^{103}$Rh, which is still being analyzed.

Not all of our measurements were made simply to test the electron-vacancy question. A few of the $\alpha_K$ values with tight uncertainties in the 2002 survey disagreed significantly from the calculations whether or not the *K* vacancy was included. These needed to be re-examined in case they revealed some serious flaw in the theory.

We have focused on high-multipolarity transitions for two reasons. First, they tend to have a longer lifetime, which makes it easier to isolate them experimentally from other competing decays; and second, the difference between $\alpha_K$ values calculated with and without the vacancy increases with the multipolarity of the transition (Nica *et al.*, 2014), which makes the outcome of the measurement more definitive. Considered individually and collectively, our results produce a clear and consistent picture of the validity of calculated $\alpha_K$ values at the one-percent level.

## 2. Experimental method

For an isolated electromagnetic transition that converts in the atomic *K* shell, the observation of a *K* x ray is a signal that an electron conversion has taken place; whereas a γ ray indicates that no conversion has taken place. If both x rays and γ rays are recorded in a measurement, then the value of $\alpha_K$ is given by

$$\alpha_K \omega_K = \frac{N_K}{N_\gamma} \cdot \frac{\varepsilon_\gamma}{\varepsilon_K}, \tag{1}$$

where $\omega_K$ is the *K*-shell fluorescence yield; $N_K$ and $N_\gamma$ are the respective peak areas of the *K* x rays and the γ ray; and $\varepsilon_K$ and $\varepsilon_\gamma$ are the respective detector photopeak efficiencies. For the fluorescence yield, we use the evaluation of Schönfeld and Janssen (1996), which quotes $\omega_K$ values typically to better than ±0.5% precision.

The key to any experimental application of Eq. (1) is the precision with which the ratio of detector efficiencies is known. In our case, the HPGe detector we use to observe both γ rays and *K* x rays – simultaneously in the same spectrum – has been meticulously calibrated for efficiency to sub-percent precision, originally over an energy range from 50 to 3500 keV (Hardy *et al.*, 2002; Helmer *et al.*, 2003, 2004), but more recently extended with ±1% precision down to 22.6 keV (Nica *et al.*, 2014), the average energy of silver *K* x rays. Over the whole energy range, precisely measured data were combined with Monte Carlo calculations from the CYLTRAN code (Halbleib *et al.*, 1992) to yield a very precise and accurate detector efficiency curve.

Strictly speaking, Eq. (1) only applies to a single isolated electromagnetic transition and such a situation is rarely met in nature. Of the nine transitions we have measured, only two fully satisfy this condition: those in $^{193}$Ir (Nica *et al*., 2004) and $^{137}$Ba (Nica *et al*., 2007, 2008). As far as is known, the isomeric decay of the second excited state of $^{193}$Ir is absolutely clear of interference: it decays exclusively by a direct *M*4 transition to the ground state, which is stable. The *M*4 transition from the 662-keV isomer in $^{137}$Ba to its ground state is illustrated in the upper left panel of Fig. 1, where it can be seen to be fed by the β decay of $^{137}$Cs. While the converted transition itself goes directly to the ground state, there is a second β-decay branch from $^{137}$Cs that populates another excited state in $^{137}$Ba; however, it is five orders of magnitude weaker and can be neglected in this context.

A second category of transitions, which we could measure and analyze directly via Eq. (1), is illustrated for $^{134}$Cs (Nica *et al.*, 2007, 2008) in the upper right panel of Fig. 1. These are cases in which the converted transition of interest is in cascade with another (single) transition that is of too low an energy to convert in the atomic *K* shell. In addition to $^{134}$Cs, the transitions in $^{119}$Sn (Nica *et al.*, 2014) and $^{197}$Pt (Nica *et al.*, 2009) fall into this category. For such cases, Eq. (1) still applies but in some cases – $^{119}$Sn and $^{197}$Pt in particular – the γ-ray peak from the low-energy transition is embedded among the *K* x rays and has to be very carefully corrected for.

Forming a third category are transitions whose cascaded partner does convert in the atomic *K* shell. The case of $^{111}$Cd (Nica *et al.*, 2016) appears in the lower right panel of Fig. 1; here the 245-keV transition is pure *E*2 and has a high enough energy that its *K*-conversion coefficient is about an order of magnitude smaller than that of the *E*3 transition of interest. The former can thus be accounted for without seriously degrading the precision with which the latter can be extracted. The situation is quite similar for the case of $^{125}$Te (Nica *et al.*, 2017b).

Finally there are cases, like that of $^{127}$Te (Nica *et al.*, 2017a), which are more complicated variants of one of the categories already described. As can be seen from the bottom left panel of Fig. 1, the *M*4 electromagnetic transition from the 106-day isomer at 88.2 keV in $^{127}$Te directly populates the ground state, but the isomeric level also has a weak β-decay branch to the 7/2$^+$ first-excited state in $^{127}$I. That state in turn decays by a mixed *M*1+*E*2 electromagnetic transition that converts in the atomic *K* shell, giving rise to iodine *K* x rays, which are unresolved from the tellurium *K* x rays needed for the application of Eq. (1) to the *M*4 transition of interest. Furthermore, the β decay of the $^{127}$Te ground state, which is in secular equilibrium with the isomeric-state decay, also has a 1.2% branch to the same 7/2$^+$ state in $^{127}$I. Fortunately, these branches are weak enough that, with care, this unavoidable interference can be accounted for without serious degradation of the final uncertainty. Our most recent case study, that of the 39.8-keV *E*3 transition in $^{103}$Rh, is similarly complicated by competing transitions fed by β decay.

In all but one case, we produced the sources used in our measurements by neutron activation, either at the Oak Ridge High Flux Isotope Reactor ($^{193m}$Ir) or at the TRIGA reactor in the Nuclear Science Center at Texas A&M University ($^{103m}$Rh, $^{111m}$Cd, $^{119m}$Sn, $^{125m}$Te, $^{127m}$Te, $^{134m}$Cs and $^{197m}$Pt). The $^{137}$Cs source used for the measurement on $^{137m}$Ba was purchased; it is a common calibration source. For each activated source, sequential spectra were recorded for up to a month in order both to acquire adequate statistics and to enable decay-rate analysis useful in the identification of – and, if necessary, correction for – activation impurities interfering with x- or γ-ray peaks needed for Eq. (1).

## 3. Results and Discussion

Of the nine transitions referred to here, we have obtained final $α_K$ results on eight. These results appear in Table 1, where they are each compared with three different theoretical calculations, all made within the Dirac-Fock framework but with one that ignores the *K*-shell vacancy and two others that include it via different approximations: the "frozen orbital" approximation, in which it is assumed that the atomic orbitals have no time to rearrange after the electron's removal; and the SCF approximation, in which the final-state continuum wave function is calculated in the self-consistent field (SCF) of the ion, assuming full relaxation of the ion orbitals.

Figure 2 plots the percentage differences between experimental and calculated $α_K$ values for two of the theoretical models: the no-vacancy version and the frozen-orbital version. Our eight completed measurements, which appear in Table 1, are represented as solid circles in the figure. Fifteen years ago, before we began our series of measurements, the survey of Raman *et al.* (2002) found only six high-multipolarity transitions (*E*3 or greater), for which $α_K$ values were known to better than ±2%: These results appear as open circles in the figure. Note that we have replaced – and significantly revised – three of the 2002 results, which are consequently shown in light gray. The three remaining pre-2002 results appear in black. Thus, there are now eleven high-precision $\alpha_K$ values for high-multipolarity transitions. All but three are from our work.

It is immediately evident from Fig. 2 that the data are completely inconsistent with ICC calculations that ignore the atomic vacancy, and that they are in remarkable agreement with the calculations that account for the vacancy in the frozen-orbital approximation. The $\chi^2$ values appearing in the bottom two rows of Table 1 quantitatively confirm this conclusion and, furthermore, indicate a preference for the frozen-orbital method of dealing with the vacancy over the SCF approach. This latter preference is almost entirely due to the $^{193m}$Ir measurement – without it, the SCF $\chi^2$ would be 1.6 (see the bottom line of the table) – but that result by itself is quite definitive: The experimental $\alpha_K$ value differs from the SCF calculation by four standard deviations. This case was chosen in the first place because the 80.22-keV *M*4 transition leads to a ~4-keV *K*-conversion electron, which is the lowest energy for any available source; consequently it is the most sensitive to the theoretical handling of that electron.

Not all of our measurements were selected because they were particularly sensitive to the vacancy/no-vacancy choice in the calculations. We were motivated to measure the transitions in $^{134}$Cs and $^{197}$Pt because their previously known results disagreed with *both* types of calculation (see Fig. 2). Perhaps they reflected additional problems with the theory. It is reassuring that both these cases turned out to have been experimentally incorrect; they were not anomalous after all. There was also an added bonus: Our precision in the $^{134}$Cs case was sufficient for that result to serve as well to discriminate between the two classes of calculation.

It is noteworthy that among the eleven precisely measured $\alpha_K$ values in Fig. 2, there are eight that statistically distinguish between the vacancy and no-vacancy calculations and they all present a consistent picture that favors inclusion of the atomic vacancy in ICC calculations. All but one of these cases come from our work.

## 4. Conclusions

We have made $\alpha_K$-value measurements on three *E*3 and six *M*4 transitions spanning the ranges $45 \leq Z \leq 78$ and $103 \leq A \leq 197$. Precise results from eight of these measurements are presented here. They represent a large fraction of all world data on $\alpha_K$ values for high-multipolarity transitions that are known to better than ±2%. In total there are now eight cases of measured $\alpha_K$ values that can contribute to answering the question of whether or not the atomic vacancy must be incorporated into ICC calculations. They all point to the need to include the vacancy, a conclusion that now fits with physical expectations based on a comparison of the known vacancy lifetimes and the time taken for the electron to exit the atom. Since these cases span a wide range of *Z* and *A* values, we can safely assert that this conclusion is generally applicable.

It is also true that the data show a preference for the frozen-orbital rather than the SCF approximation being used to account for the vacancy, but this is almost entirely based on one particularly sensitive case: that of the decay of $^{193m}$Ir. Consequently we cannot pronounce the general validity of this preference, although we suspect it to be true.

Early results from our program influenced a reevaluation of ICCs by Kibédi *et al.* (2008) and the development of BrIcc, a new data-base obtained using the code by Band *et al.* (1993), which is based on Dirac-Fock calculations. In conformity with our conclusions, a version of this code was used, which incorporates the "frozen orbital" approximation to account for the atomic hole. The BrIcc data-base is available on-line for the determination of ICCs.[†] It has also been adopted by the National Nuclear Data Center (NNDC) and by the Decay Data Evaluation Project (DDEP). Obviously our more recent results have continued to support that decision by showing that the same conclusion applies to transitions in nuclei with a widening range of $Z$ values.

## Acknowledgements

This work is supported by the U.S. Department of Energy under Grant No. DE-FG02-93ER40773 and by the Robert A. Welch Foundation under Grant No. A-1397.

## Footnotes

* Corresponding author. Tel: +01-979-845-1411; fax: +01-979-845-1899.

  E-mail address: hardy@comp.tamu.edu

[†] Available at http://bricc.anu.edu.au

**Table 1:** Measured $\alpha_K$ values for eight of the precisely measured transitions considered in this work. The experimental results are compared with calculated values that were based on three different assumptions as described in the text: 1) the atomic vacancy is ignored; 2) the vacancy is included but the other atomic orbitals of the initial-state atom are considered "frozen"; and 3) the vacancy is included but the other orbitals satisfy the self-consistent field (SCF) of the final-state ion. The uncertainties on the calculated values caused by uncertainties on the transition energies are omitted from the table as being negligible by comparison to the experimental uncertainties.

| Parent state | Transition Energy (keV) | Measured $\alpha_K$ | Calculated $\alpha_K$ values based on following assumptions | | |
|---|---|---|---|---|---|
| | | | No vacancy | "frozen orbitals" | SCF |
| $^{111m}$Cd | 150.825(15) | 1.449(18) | 1.425 | 1.451 | 1.446 |
| $^{134m}$Cs | 127.502(3) | 2.742(15) | 2.677 | 2.741 | 2.730 |
| $^{119m}$Sn | 65.660(10) | 1621(25) | 1544 | 1618 | 1603 |
| $^{125m}$Te | 109.276(15) | 185.0(40) | 179.5 | 185.2 | 184.2 |
| $^{127m}$Te | 88.23(7) | 484(6) | 468.6 | 486.4 | 483.1 |
| $^{137m}$Ba | 661.659(3) | 0.0915(5) | 0.09068 | 0.09148 | 0.09139 |
| $^{193m}$Ir | 80.22(2) | 103.0(8) | 92.0 | 103.3 | 99.7 |
| $^{197m}$Pt | 346.5(2) | 4.23(7) | 4.191 | 4.276 | 4.265 |
| $\chi^2$ | | | 231 | 0.77 | 18.6 |
| $\chi^2$ (without $^{193m}$Ir) | | | 41.5 | 0.63 | 1.6 |

**Figure 1**: Partial decay schemes relevant to our measurements of $\alpha_K$ values. The top two panels are taken from Nica *et al.* (2007); the bottom left is from Nica *et al.* (2016); and the bottom right is from Nica *et al.* (2017a).

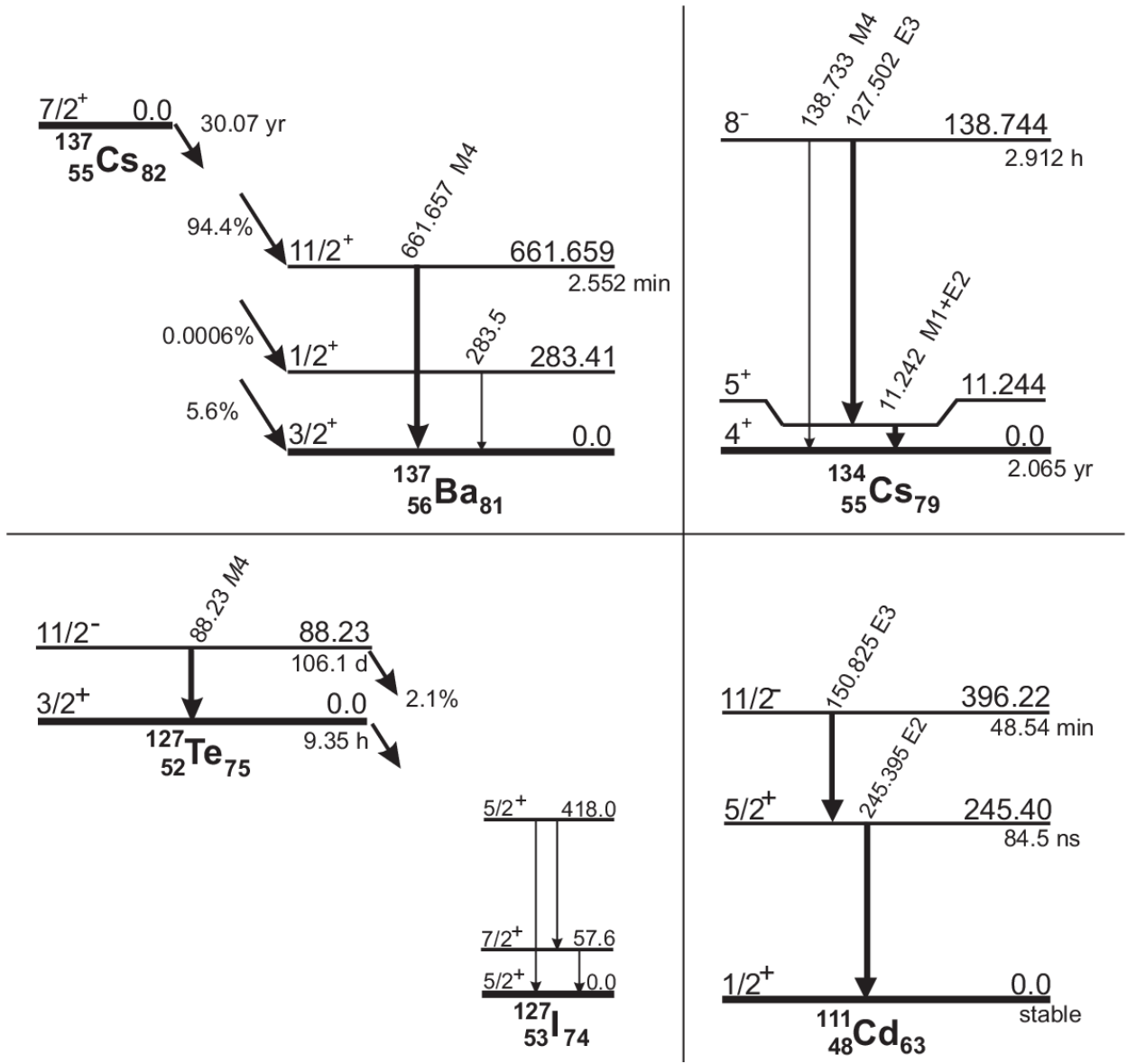

**Figure 2:** Percentage differences between the measured and calculated $\alpha_K$ values – expressed as (experiment-theory)/theory – for two Dirac-Fock calculations: one (a) is without any provision for the atomic vacancy and the other (b) is with the vacancy included according to the "frozen orbital" approximation. Solid (red) circles are our measurements; open circles refer to pre-2002 results, the ones in gray having been replaced. The figure shows all $\alpha_K$ values for high-multipolarity transitions (*E*3 and above) that are known to better than ±2%.

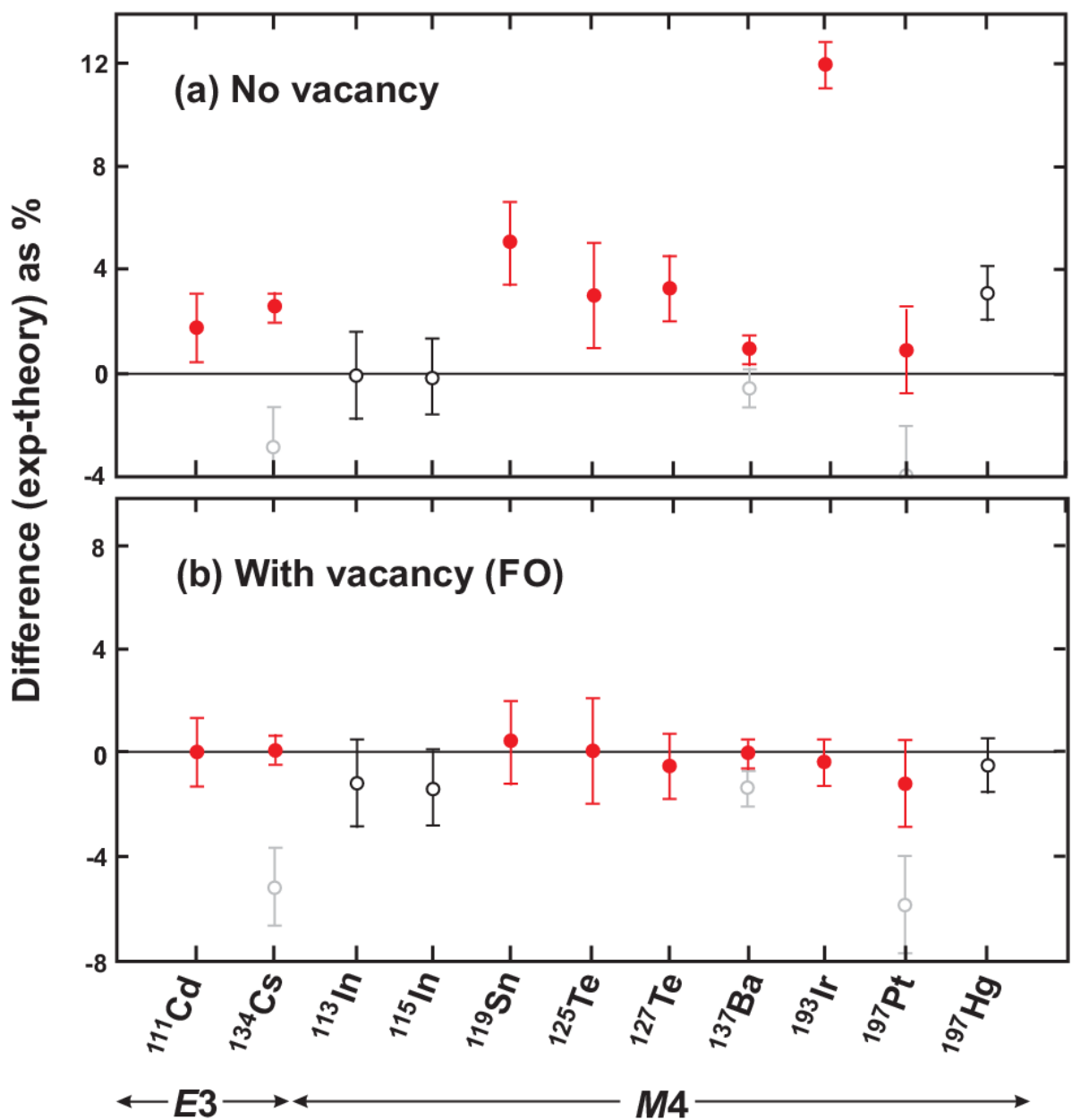